\documentclass[preprintnumbers,amsmath,amssymb]{revtex4}
\usepackage{graphicx}% Include figure files
\usepackage{dcolumn}
\usepackage{bm}% bold math
\def\eqnn#1{(\ref{eq:#1})}
\def\plb{{ \sl Phys. Lett. }}

\def\jsp{{\sl  J. Stat. Phys.}}
\def\figno#1{Fig.~\ref{fig:#1}}
\def\cum#1{\langle\langle#1\rangle\rangle}
\def\vev#1{\langle#1\rangle}
\begin{document}
\title{Spatial Distributions of Observables in Systems under Thermal
Gradients}
\author{
  Kenichiro Aoki\footnote{E--mail:~{\tt ken@phys-h.keio.ac.jp}.
      Supported in part by the Grant--in--Aid from
      the Ministry of Education, Science, Sports and Culture.}
  and Dimitri Kusnezov\footnote{E--mail: {\tt
      dimitri@mirage.physics.yale.edu} }}
\affiliation{$^*$Dept. of Physics, Keio University, {\it
    4---1---1} Hiyoshi, Kouhoku--ku, Yokohama 223--8521, Japan\\
  $^\dagger$Center for Theoretical Physics, Sloane Physics Lab, Yale
  University, New Haven, CT\ 06520-8120} \date{\today }

\begin{abstract}
  Departures of observables from their thermal equilibrium expectation
  values are studied under heat flow in steady-state non-equilibrium
  environments. The relation between the spatial and temperature
  dependence of these   non-equilibrium  behaviors and the underlying
  statistical properties are clarified from general
  considerations. The predictions 
  are then confirmed in direct numerical simulations within the
  FPU-$\beta$ model. Non-equilibrium momentum distribution functions
  are also examined and characterized through their cumulants and the
  properties of higher order cumulants are discussed.
  \end{abstract}

\vspace{3mm}

% 05.60.Cd: classical transport
% 63.10.+a: General lattice theory
% 44.10.+i: heat conduction
% 11.10.Wx: finite T field theory
\maketitle
\section{Introduction}
In studies of non-equilibrium physics, especially those of steady
states, local equilibrium is most often invoked and this assumption
simplifies calculations through the use of equilibrium statistical
mechanics and thermodynamics\cite{neq}. The local equilibrium
assumption allows the use of the equilibrium distribution function to
compute observables. If local equilibrium conditions are not assumed,
very little can be computed analytically and even the definition of
temperature is 
no longer unique\cite{keizer,nonEqT}.  Efforts have been made to
quantify the goodness of local equilibrium assumptions or how
transport coefficients differ from their linear response values,
though only few quantitative studies
exist\cite{fluct,eos,hk,local-eq,onsager,dhar,takesue,ak-le}.  Without
the knowledge of the non-equilibrium steady-state distribution,
theoretical development becomes quite restrictive.  We explore how
observables depart from their equilibrium expectation values within a
given non-equilibrium steady-state, specifically focusing on the
spatial dependence of the non-equilibrium expectation values within a
given system and their local temperature dependence. To make this
concrete, heat flow in the FPU $\beta-$model is simulated to test the
predictions. We further quantitatively examine the relationship
between  the momentum cumulants and the distribution and find that the
lower order cumulants characterize the distribution quite well.

For systems in thermal gradients, it is natural to consider how an
observable $\cal O$ in the non-equilibrium steady state departs
from its equilibrium value, denoted ${\cal O}_{eq}$. The
normalized deviation from equilibrium, when  ${\cal
O}_{eq}\not=0$, can be expanded as
\begin{equation}
  \label{eq:le1}
  \delta_{\cal O} \equiv\frac{\delta {\cal O}}{\cal O}=\frac{{\cal
      O}-{\cal O}_{eq}}{{\cal O}_{eq}} =
  C_{\cal O}\left[\frac{\nabla T}{T}\right]^2 + C_{\cal O}'\left[\frac{\nabla
      T}{T}\right]^4 + \cdots
\end{equation}
When ${\cal O}_{eq}=0$, as is the case for higher order momentum
cumulants, one can normalize by an observable which has
the same dimensions.
When local equilibrium is no longer valid, in general, no unique
definition of temperature exists and a choice needs to be made.  This
definition of non-equilibrium temperature can be thought of as a
choice of a coordinate system, on which the physics behavior of the
system will not depend.
If we assume analyticity in $\nabla T$, the deviations $\delta_{\cal
  O}$ can be expanded
in even powers as above.  We shall see below that this expansion is
adequate for describing the properties of the system.

The heat flow, $J$, is the flow of energy and can be unambiguously
defined in Hamiltonian systems. Near equilibrium, it satisfies
Fourier's law locally as $J=-\kappa \nabla T(x)$, where $\kappa$ is
the thermal conductivity, $T(x)$ is the temperature profile inside,
and $x$ is the position inside the system. Fourier's law can be used
in \eqnn{le1} to re-express the local departures from equilibrium in
terms of the temperature profile $T(x)$, or equivalently the position
$x$ once the coefficients $C$, $C'$ are known since $J$ does {\it not}
depend on $x$;
\begin{equation}
  \label{eq:spatial1}
  \delta_{\cal O} =
  C_{\cal O}\left(J\over\kappa(T) T\right)^{2}+
  D_{\cal O}' \left(J\over\kappa(T) T\right)^{4}+  \ldots
\end{equation}
We note that Fourier's law itself receives non-equilibrium
corrections\cite{ak-le}, which is why the coefficient of ${\cal
  O}(J^4)$ term in the expansion \eqnn{spatial1} differs from that of
\eqnn{le1}.  In the following, the objectives will be to  make the
formula more explicit and understand its physical properties under
rather general assumptions. This relation, together with $\kappa(T)$
(and consequently $T(x)$) provides the basis for defining how
non-equilibrium observables vary inside a finite system both near and
far from global thermal equilibrium.
\section{The FPU model and Temperature Profiles}
The results we present here are derived from general
considerations and we develop them in conjunction with a model in
which they can be explicitly analyzed. We study the FPU $\beta$
Hamiltonian, defined generally in the form
\begin{equation}\label{eq:fpu-dimensionful}
{\tilde H}  = \sum_{k=0}^L \left[
  \frac{\tilde p_k^2}{2m}+\frac{1}{2}m\omega^2
 (\tilde q_{k+1}-\tilde q_k)^2 + \frac{\beta}{4} (\tilde
  q_{k+1}-\tilde q_k)^4\right].
\end{equation}
We use the FPU model since its physical properties are of wide
interest (\cite{km,lepri,ak-fpu,fpuRevs} and references therein).
Also as the model is well studied, we can understand the physical
properties we find within a larger physics context.
Under the
rescaling $\tilde p_k=p'_k\omega^2\sqrt{m^3}$, $\tilde q_k=
q'_k\omega\sqrt{m}$, we obtain the conventional form of the
FPU~$\beta$ model,
\begin{equation}\label{eq:fpu-usual}
 H_\beta = \frac{1}{2}\sum_{k=0}^L \left[ p_k^{\prime 2}
  + (q'_{k+1}- q'_k)^2
  + \frac{\beta}{2} (q'_{k+1}- q'_k)^4\right],
\end{equation}
where $H_\beta=\tilde H/(m^2\omega^4)$. We note that in finite
temperature simulations, changing the temperature is equivalent to
changing the coupling $\beta$. Under the additional rescaling
$p'_k=p_k/\sqrt\beta $, $q'_k=q_k/\sqrt\beta$, one obtains a
unique, dimensionless, Hamiltonian $H\equiv H_{\beta=1}=\beta
H_\beta$, which we shall use without any loss of generality. Since
$p_k^2=\beta p_k'^2$, the temperatures in the two formulations $H$
and $H_\beta$ are related by $T=\beta T'$.

In this work, we study the non-equilibrium steady state physics of the
theory under thermal gradients, making use of non-equilibrium states
constructed numerically.  (For general discussion, see, for instance,
\cite{neqRevs,thermoRevs}.) The model is thermostatted at the
boundaries $k=0,L$ at various temperatures $T_1^0, T_2^0$, using the
generalized versions of Nos\'e--Hoover thermostats as detailed in
\cite{ak-long}.  These additional thermostat degrees of freedom are
added only at the boundaries and the degrees of freedom inside the
system ($0<k<L$) are exclusively those of the Hamiltonian
\eqnn{fpu-usual}.  By numerically integrating the equations of motion
of the whole system (including those of the thermostats), we obtain
the  behavior of physical observables in the non--equilibrium steady
state by averaging over time, in the standard manner\cite{thermoRevs}.
The local temperature at site $k$ is defined as $T_k=\langle
p_k^2\rangle$.  In this work, we study the physics inside the system,
away from the boundaries by much more than the mean free path of the
system\cite{ak-fpu}.  The sensitivity of the results to the manner in
which we apply the boundary conditions --- including both the number
of thermostats and the strength of the couplings --- have been
examined to ensure that physics results below remain independent of
their implementation. (The only exceptions are the boundary jumps in
temperature which we discuss below.) The numerical integrations were
performed using the fourth order Runge-Kutta routines with time steps
of $0.005\sim0.02$ for $10^7\sim10^{10}$ time steps.  The equilibrium
properties have been readily verified with this
method\cite{ak-fpu,ak-long}.

\begin{figure}[hptb]
  \centering
  \includegraphics[width=8.5cm,clip=true]{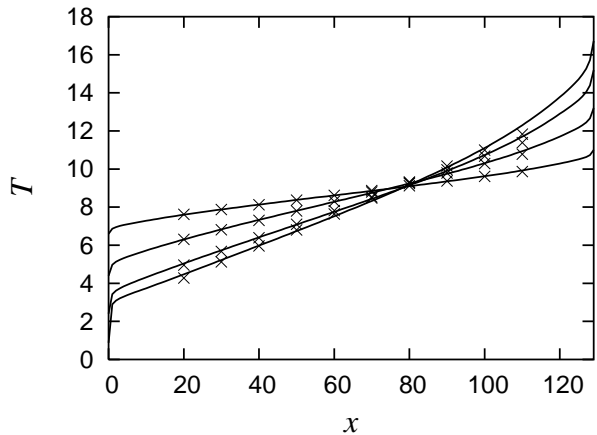}
  \caption{Some examples of temperature profiles for the FPU model
  with $L=128$.
  The thermostat temperatures at the boundaries are
  $(T_1^0,T_2^0)=(0.88,16.72),(2.4,15.2),(4.4,13.2),(6.6,11.0)$ for
  the four thermal profiles.  The profiles predicted from Eq.~\eqnn{t-profile}
  are indicated by $\times$ and agree well with the
  results from the numerical simulations.}
  \label{fig:p2x}
\end{figure}
In \figno{p2x} some examples of temperature profiles for the FPU
theory are shown.  Generically, there are temperature jumps just inside
the boundaries with smooth temperature variations within. The boundary
jumps become larger as one moves away from global equilibrium. The
jumps are dynamical in the sense that they depend on the model,
the transport
coefficient, heat flow, as well as the type of boundary conditions
employed. The temperatures at the boundaries are at the thermostat
temperatures to high degree of precision. For instance, in the
examples of \figno{p2x}, the boundary temperatures are equal to
the prescribed thermostat temperatures to within few in $10^5$
relatively.

From temperature profiles and heat flow calculations,
Fourier's law can be verified to hold
up to corrections of the form \eqnn{le1}, and the thermal
conductivity, $\kappa$, can be obtained for a given temperature and
system size. In the 1-d FPU model, $\kappa$ depends on the system
size $L$ and does not display bulk behavior\cite{lepri}. $\kappa$
is also dependent on the temperature in a known
manner\cite{ak-fpu}. Generally, in cases where we have a one
dimensional temperature gradient, the temperature profiles can be
obtained by integrating Fourier's law
as long as we are not too far from equilibrium
\cite{ak-le,profs,ak-long}:
\begin{equation}
  \label{eq:integrateFL}
  \int_{T_1}^{T(x)} \kappa(T)\,dT=-Jx,\qquad
  J=-p_k\left[(q_{k+1}-q_k)+(q_{k+1}-q_k)^3\right]
\end{equation}
$x$ is the continuum extrapolation of the discrete lattice index $k$.
We note here that $J$ is a constant within the system for a given set
of temperature boundary conditions since there are no heat sinks or
sources inside. $T_1$ in the integral is the temperature extrapolated
to the boundary and is explained below.

In many situations, the temperature dependence of the thermal
conductivity, within some temperature range, can be well described
by
\begin{equation}
  \kappa(T) = cT^{-\gamma}.
  \label{eq:tc}
\end{equation}
While this power law  may not hold globally in $T$, it is often the case that
it is sufficient for the region of interest, which is the case
here. In such a situation, the temperature profile can be explicitly
computed from \eqnn{integrateFL} to be\cite{ak-long}
\begin{equation} \label{eq:t-profile}
  T(x) = \left\{\begin{array}{ll}
    T_1\left[1-\left(1-\left(\frac{T_2}{T_1} \right)^{1-\gamma}
      \right)
      {\frac{x}{L}}\right]^{{\frac{1}{1-\gamma}}},\qquad &
    \gamma\neq1\\
    T_1 \left( \frac{T_2}{T_1}\right)^{x/L}, &\gamma=1\quad.
\end{array}\right.
\end{equation}
Here, $T_{1,2}$ denote the boundary temperatures obtained by
extrapolating the temperature profile inside the system and
differs from the thermostat temperatures $T_{1,2}^0$ by the
boundary temperature jumps. From \eqnn{integrateFL} and \eqnn{tc},
the temperatures $T_{1,2}$ are found to obey a relation
\begin{equation}\label{eq:one}
  -{ JL\over c}={ T_2^{1-\gamma} -  T_1^{1-\gamma}\over1-\gamma}
\end{equation}

To understand the temperature profile of the whole system, we
further need an understanding of the temperature jumps at the
boundaries\cite{ak-jumps}. Similar boundary slips have been seen
in sheared systems and these effects have been known for a long
time in real systems.  To leading order, the temperature jumps can
be described by (with $n$ being the normal to the boundary)
\begin{eqnarray}
  \left|T_i - T_i^0\right|
  \backsimeq \frac{\alpha c}{L(1-\gamma)}\left[
    T_2^{1-\gamma}-T_1^{1-\gamma}\right]
  \sim \lambda\frac{\partial T}{\partial n},\qquad (i=1,2)
  \label{eq:jumps}
\end{eqnarray}
Here $\lambda$ is the mean free path of the excitations, which for
the FPU lattice model, is essentially the $\kappa(T)$ (up to a
constant factor of order one) due to kinetic theory
arguments\cite{ak-fpu}.
$\alpha$ reflects the efficacy of the boundary conditions. The
last relation is obtained by using Fourier's Law and \eqnn{one}.
The jumps on the hot and cold side are the same provided the
system is reasonably close to equilibrium. The jumps at the
boundaries and the temperature profile within \eqnn{t-profile}
describe the temperature profile of the complete system.
The predicted values for the temperature profiles are plotted in
\figno{p2x} at a number points inside the systems ($\times$
symbols) away from the boundaries and are seen to be consistent
with the simulation results. The thermal conductivity is roughly
constant with respect to the temperature in this region so that
$\gamma=0$ was used in the profile calculations. This demonstrates
that all aspects of the non-equilibrium temperature profile can be
quantitatively captured through \eqnn{t-profile} and \eqnn{jumps},
irrespective of whether
$\kappa(T)$ is a power law in temperature for all $T$ or not. With
this understanding of $T(x)$ we can now turn to the question of
general observables.
\section{Spatial Dependence of Cumulants in the  Non-equilibrium
  Steady State}
In non-equilibrium steady states, physical observables show
deviations from their equilibrium values reflecting the lack of
local equilibrium in the system. The behavior of the observables
have been seen to be  well described by \eqnn{le1} on average, at
least in some cases\cite{ak-le}. Here,  we now would like to
investigate a more detailed issue --- whether these properties can be
used to understand the nature of the spatial profiles of these
observables in a given non-equilibrium situation. We will assume that
within some range of $T$ and $L$ that we can represent the expansion
coefficients in \eqnn{le1} as
\begin{equation}\label{eq:Cbehavior}
  C_{\cal O}=\mu_{\cal O} T^{s_{\cal O}} L^{\alpha_{\cal O}}
\end{equation}
The behavior of $C_{\cal O}$ with respect to $T,L$ clearly must depend
on the dynamics of the theory and is not expected to be generic.
\begin{table}
\begin{tabular}{lrr}
\hline
 FPU $\beta-$ Model in $d=1$:   &\quad $(\mu T^s)$
  & $\alpha$   \\
\hline
& &\\
$T=1$ & 29(5) & 0.87(4)\\
$T=8.8$ & 13(1)   & \quad 0.99(1) \\
$T=88$ & 7.4(4)   & 1.04(2) \\
\hline
\end{tabular}
\qquad\qquad
\begin{tabular}{rrr}\hline
$\phi^4$ Theory :
   & \qquad $(\mu T^s)$
   & $\alpha$   \\\hline
& & \\
$d=1\quad T=1$ & 3.3(24) & 0.96(15)\\
$T=5$ & 1.6(6)   & 1.18(9) \\
$d=2\quad T=1$ & 1.9(4)   & 1.09(5) \\
$T=5$ & 0.4(2)   & 1.6(2)  \\
$d=3\quad T=1$ & 4 (1)       &\ \  0.96(10)    \\
$T=5$ & 0.2(5)   & 1.6(6)  \\
\hline
\end{tabular}
\caption{Non-Equilibrium coefficients $C_{4}=(\mu T^s) L^\alpha$
for
  $\cum{p^4}/T^2$ ({\it cf.} Eq.~\eqnn{Cbehavior},\eqnn{fpuT}).
  The results are shown for the FPU $\beta$ model and the $\phi^4$
  theory in   $d=1\sim3$--dimensions.
The value of $s$ is extracted from fitting to several
temperatures. } \label{tab:a}
\end{table}

To study the spatial distribution of physical observables in
non-equilibrium, we make use of \eqnn{spatial1} which describes
how the observables should behave in non-equilibrium locally in
space, given the thermal conductivity. Using this property and
\eqnn{tc}, we obtain to leading order that observables will
deviate from their local equilibrium values as
\begin{equation}\label{eq:dist}
 \delta_{\cal O} =  C_{\cal O}\; \left(\frac{J
 T(x)^{\gamma-1}}{c}\right)^2  = a_{\cal O}T(x)^{2(\gamma-1)+s}
\end{equation}
Here $a_{\cal O}$ is defined through this equation and should be
proportional to $J^2$. This implicitly contains the spatial
distribution since the temperature profile is known  and can be
understood as in \eqnn{t-profile}.

While these arguments apply to any physical observable in the
system, we choose to study cumulants of momenta, $p$, mainly for
the following reasons; conceptual and practical.  There seems to
be no universal rigorous definition of local equilibrium, yet the
concept in the least seems to include a unique meaning for
temperature, which in this case would lead to the Maxwellian
distribution for $p$. To put another way, when the momentum
distribution is not Maxwellian, we can choose different definitions
of the temperature based on the various moments of
$p$\cite{keizer,nonEqT}. The
cumulants of the momentum distribution provide insight into how the
physical properties of a non-equilibrium system deviates from those of
local equilibrium.
The cumulants are well defined local variables and their values in
local equilibrium are known precisely. The low order cumulants
are defined as
\begin{equation}
  \cum{p^2}=\vev{p^2},\quad
  \cum{p^4}=\vev{p^4}-3\vev{p^2}^2,\quad
  \cum{p^6}=\vev{p^6}-15\vev{p^2}\vev{p^4}+30\vev{p^2}^3  ,\ \ldots
\end{equation}
where, in {\it equilibrium},
\begin{equation}
  \label{eq:cumValues}
  \cum{p^2}_{eq}=T,\qquad\cum{p^n}_{eq}=0\ (n\not=2)
\end{equation}
This property is also of practical importance. Since the deviations we
compute can be small, it is desirable to use observables
whose local equilibrium values are  known exactly.  In this case
in thermal equilibrium, ${\cal O}_{eq}=0$, so we use $\delta_{\cal
O} = \cum{p^{2n}}/T^n$. We list the coefficient for the case
${\cal O}=\cum{p^4}$ in Table~\ref{tab:a} for the FPU
$\beta-$model as well as $\phi^4$ theory\cite{ak-long} for
comparison.

\begin{figure}
\includegraphics[width=8.5cm,clip=true]{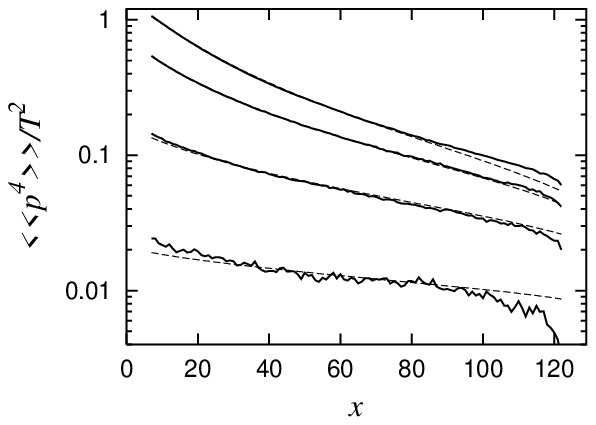}
\includegraphics[width=8.5cm,clip=true]{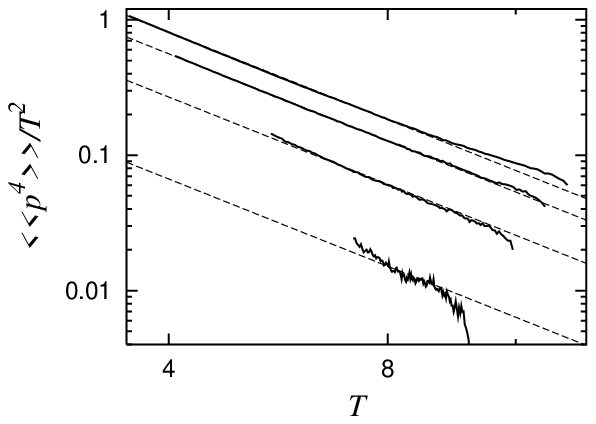}
\caption{(left)  Spatial dependence of the rescaled 4-th momentum
  cumulant, $\cum{p^4}/T^2$ for the four systems in \figno{p2x}.
  Larger cumulant values are seen for larger boundary temperature
  differences.   (right) Temperature dependence of $\cum{p^4}/T^2$ for the same
  systems. In both panels, the predictions \eqnn{dist} are indicated by
  the dashes.
}
\label{fig:p4}
\end{figure}

\begin{figure}
\includegraphics[width=8.5cm,clip=true]{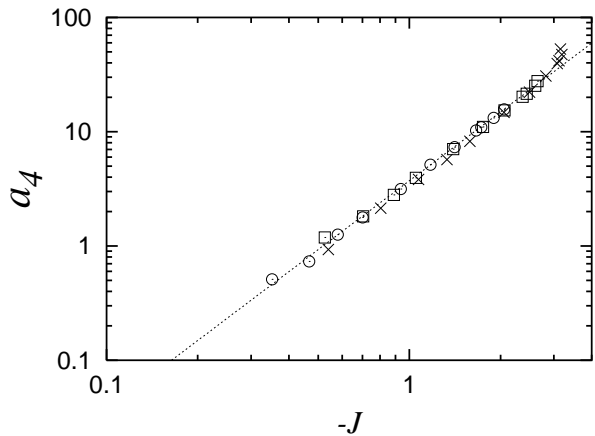}
\caption{\label{fig:a2}  $J$ dependence of the non-equilibrium
  expansion coefficient, $a_4$,   for various boundary conditions,
  $(T_1^0,T_2^0)$  and system sizes $L$. The dashed line is $3.72\,J^2$
  and the $\sim J^2$ behavior of the coefficient can be clearly seen,
  as predicted from theory. Each data point represents a
  particular temperature boundary condition for
  $L=32$ ($\times$), $L=64$ ($\Box$) and $L=128$ ($\bigcirc$) systems.}
\end{figure}

Let us investigate how well \eqnn{dist} describes the spatial
distribution of $\cum{p^4}/T^2$. We find
\begin{equation}
 {\cum{p^4}\over T^2}
 =a_4T^{2(\gamma-1)+s_4}
 = a_4\left(T_1\left[1-
     \left(1-\left(\frac{T_2}{T_1} \right)^{1-\gamma} \right)
     {\frac{x}{L}}\right]^{{\frac{1}{1-\gamma}}}\right)^{2(\gamma-1)+s_4}
  \label{eq:fpuT}
\end{equation}
$s_4$ is the temperature dependence of the coefficient $C_4$ which is
reflected in Table~\ref{tab:a}. To understand the validity of the
prediction Eq.~\eqnn{fpuT}, fits were made with just one parameter
$a_4$ for the whole profile. We find that this describes the situation
quite well, as seen in the examples of \figno{p4}, where the
predictions are denoted by dashes.  In these figures, we have compared
the fits with the spatial as well temperature dependence of
$\cum{p^4}$ for the four systems shown in \figno{p2x}. In this
temperature range, temperature dependence of the thermal conductivity
is weak so we used $\gamma=0$ and $s_4=-0.14$ extracted from
the data in Table~\ref{tab:a}.  Similar results were found for
different temperature boundary conditions and for different $L$.
To further verify the underlying physics, we study the $J$ dependence
of the coefficient $a_4$.
The behavior for various systems, including the four systems in
\figno{p2x}, are shown in \figno{a2}. Each data point represents a
system with a particular size and temperature boundary conditions. The
central temperature is around $T=8.8$  and is kept fixed.
The observed behavior is clearly well described by $a_4\sim$
const.$\times J^2$.  The coefficient $a_4$ seems $L$-independent and
this can roughly be understood since $c^2$ grows in $L$ in a manner
similar to $C_{4}$.  We have in addition systematically studied the
results to see if we can discern the contribution of higher order
terms in the expansions \eqnn{le1},\eqnn{spatial1} (of order $J^4$ and
higher)  but have found no consistent evidence for them.
In other physical situations, non-analytic behavior seems to have been
seen in some cases\cite{shears,sasa}.

\begin{figure}
\includegraphics[width=8.5cm,clip=true]{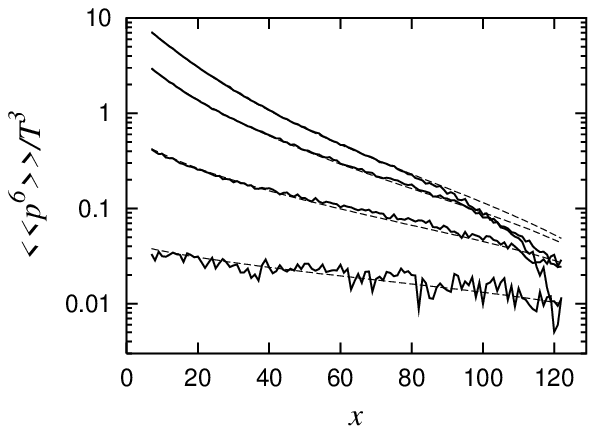}
\includegraphics[width=8.5cm,clip=true]{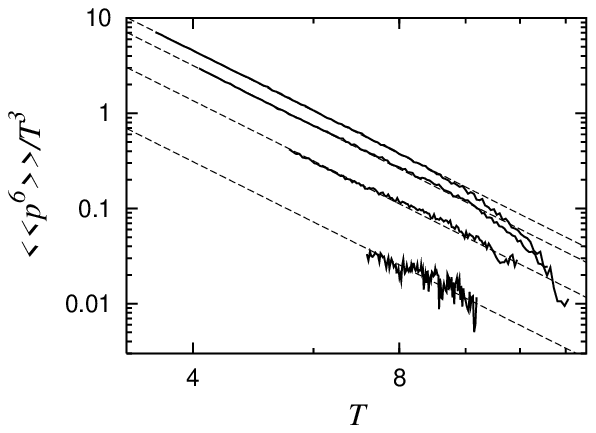}
\caption{Spatial dependence (left) and temperature dependence
  (right) of $\cum{p^6}/T^3$ for the four systems in \figno{p2x}.
  Larger cumulant values are seen for larger boundary temperature
  differences. Predictions are shown with dashes. }
\label{fig:p6}
\end{figure}

\begin{figure}
\includegraphics[width=8.5cm,clip=true]{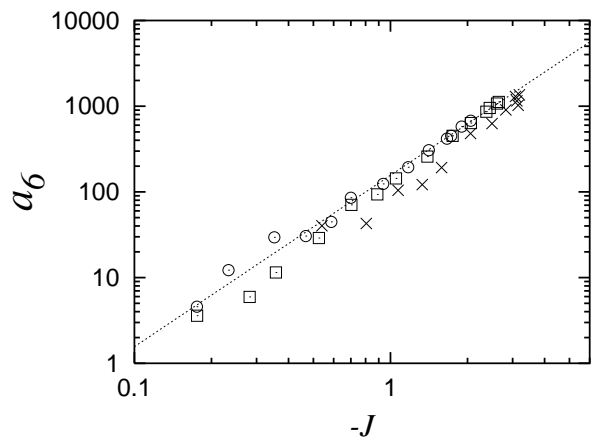}
\caption{\label{fig:p63}  $J$ dependence of the coefficient $a_6$ for
  various boundary conditions, $(T_1^0,T_2^0)$  and system sizes $L$.
  The dashed line denotes  $156\, J^2$. $\sim J^2$dependence of $a_6$
  is evident, 
  in agreement with the predictions.  Each data point represents a
  particular temperature boundary condition for
  system sizes $L=32$ ($\times$), $L=64$ ($\Box$) and $L=128$
  ($\bigcirc$), as in \figno{a2}.}
\end{figure}

While the logic seems to work for the lowest non-trivial order
cumulant, $\cum{p^4}$, we find it instructive to analyze if it works
at higher orders. In this direction, we have analyzed
the next non-trivial order $\cum{p^6}$ and have found that its
behavior is quite consistent with  physics of \eqnn{dist}, as was the
case of $\cum{p^4}$, in all the systems we have studied.  In practice,
higher order cumulants are more prone to errors and the computations
are more difficult.  The results for the same four systems in
\figno{p2x} are shown in \figno{p6}.
As in the $\cum{p^4}$ case, the coefficient $a_6$ shows $J^2$ behavior
within error, as it should. $a_6$ shows a weak $L$ dependence, as we
would generically expect. A common value of $s_6=-1.6$ was adopted for
all the data in \figno{p6} and \figno{p63}. What is evident is that
the spatial behavior of non-equilibrium observables can be explicitly
related to transport and other physical properties of the system
using rather general considerations. From the
cumulants we now consider what can be said about the full momentum
distribution function.
\section{Cumulants and the distribution}
\label{sec:distro}
The cumulants are quantitative indicators of the non--Maxwellian
nature of the momentum distribution or the violations of local
equilibrium.  All the cumulants are non-zero unless the system is in
local equilibrium, in which case only the linear and quadratic
cumulants are non-zero.
There are very few problems where cumulants can all
be computed analytically and it becomes numerically intractable
to compute them as we go to higher orders. 
It is then of interest to see how well the lower order
cumulants characterize the distribution. The cumulants are properties
of the distribution function, which has an infinite number of degrees of 
freedom. A priori, there is no reason to assume that the lower order
cumulants characterize the distribution.  In order to clarify this
issue, first note that the distribution function $f(p)$ and the
cumulants are related explicitly through the generating function as
\begin{equation}
  \label{eq:genfn}
  \int dp\,e^{iup} f(p)= \vev{e^{iup}}
  =\exp\left(\sum_{n=0}^\infty{i^nu^n\over n!}\cum{p^n}\right)
  =\exp\left(\sum_{n=0}^\infty{(-u^2)^n\over (2n)!}\cum{p^{2n}}\right)
\end{equation}
Here, in the last equality, the symmetry under $p\leftrightarrow-p$
was used, which leads to $\cum{p^{2n+1}}=0$. We see from this equation
that  given all the cumulants (or equivalently, moments), we may recover
the distribution function by performing an inverse Fourier
transform.
However, in practice, not all the
cumulants are available. 

\begin{figure}
  \includegraphics[width=8.5cm,clip=true]{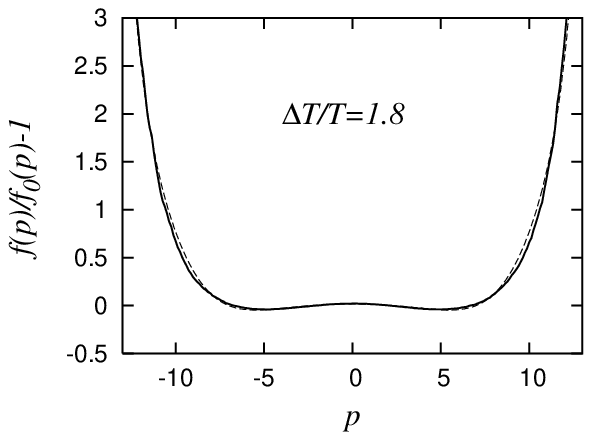}
  \includegraphics[width=8.5cm,clip=true]{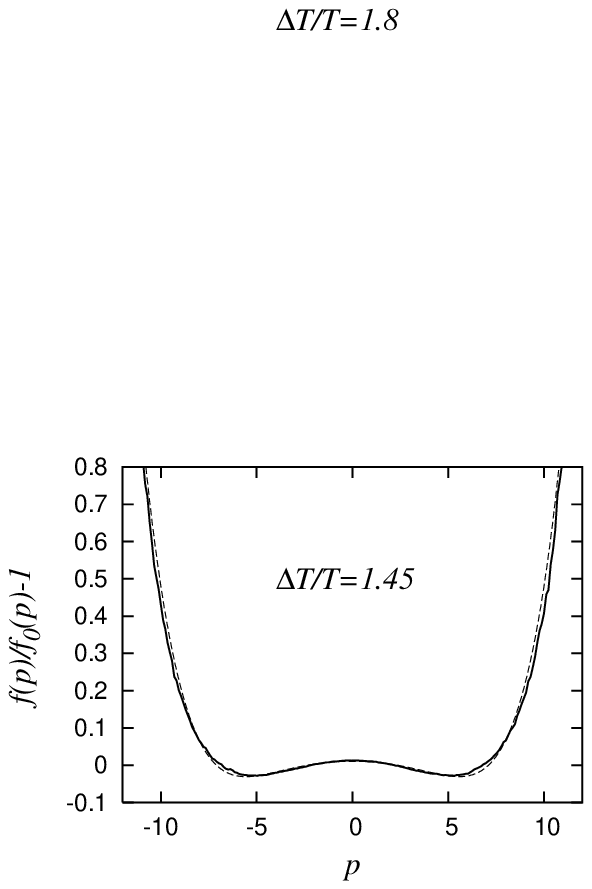}
  \includegraphics[width=8.5cm,clip=true]{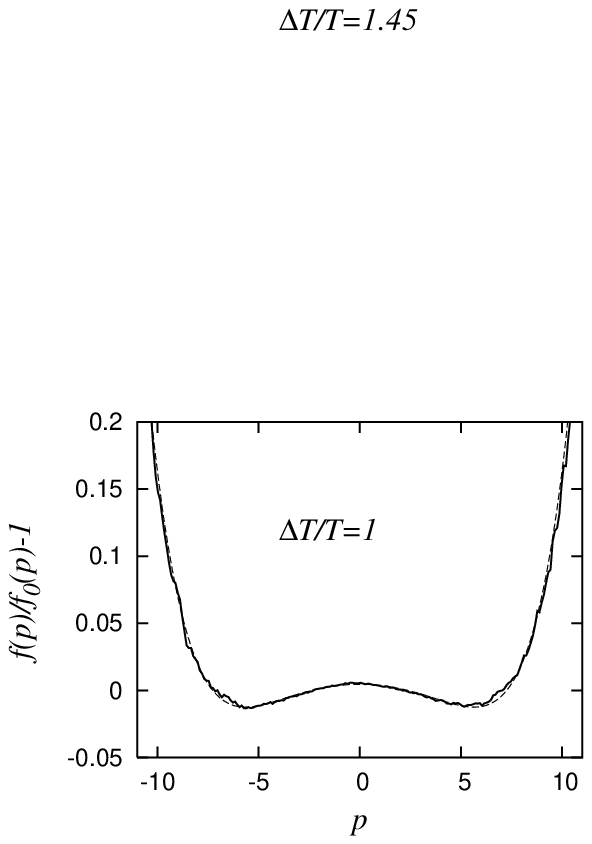}
  \includegraphics[width=8.5cm,clip=true]{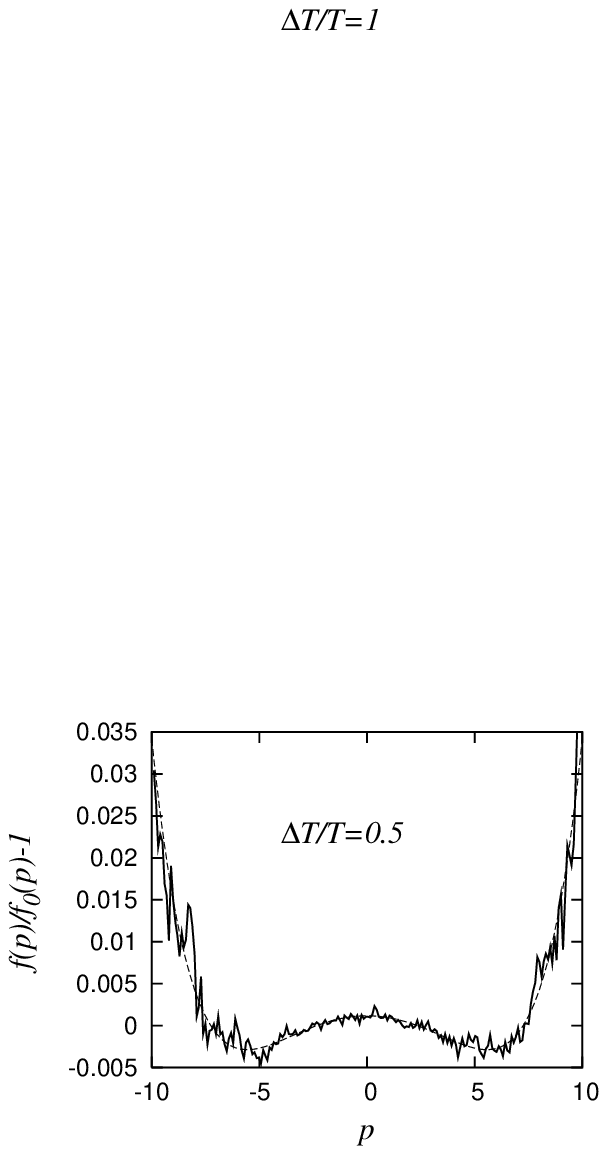}
\caption{The relative deviation of the distribution from the Maxwell
  distribution  for the four systems in \figno{p2x}. Distribution
  obtained from the cumulants $\cum{p^4},\cum{p^6}$ (dashed) are
  compared with the measured distributions (solid).
  The agreement is   excellent.
  $\Delta T/T$ denotes the boundary temperature difference over the
  average temperature and is an indication of how far the system is
  from equilibrium.
}
\label{fig:pDist}
\end{figure}

Intuitively, we expect the lower order
cumulants to be the leading order results with higher order cumulants
becoming more important as we move further away from equilibrium.
In \figno{pDist},  we plot the {\it relative difference} of the
measured distribution $f(p)$ to the thermal distribution, $f_0(p)$,
for the distribution directly measured in the simulations and the
distribution computed from the low order cumulants, $\cum{p^{2,4,6}}$.
The comparisons are performed for the four systems in \figno{p2x} at a
point in the middle of the system. 
From these graphs, we observe the following: (a) The agreement between
the distribution computed from lower order cumulants and the
distribution is quite good in all cases; (b) the relative deviation
from the thermal distribution is larger as we move away from
equilibrium (larger $\Delta T/T$), as expected; (c) the small
discrepancy between the computed distribution and the measured one
seems to be larger for larger $\Delta T/T$; (d)  the deviation from
the thermal distribution becomes more noisy for smaller $\Delta T/T$,
since the deviation itself is smaller and the relative error is
larger. We mention here that strictly speaking, the distributions can
have different behavior, such as long tails, beyond the region we have
investigated. However, these 
tails would have to be quite small since the distributions 
decay as $\exp(-p^2/(2T))$ and the agreement is good up to
reasonably large $p$, as seen in \figno{pDist}.  We have examined
numerous systems
for different $T$ and $L$ and found similar good agreement. %
Therefore, we see that the lower order cumulants provide good physical
observables  that quantitatively describe the deviations of the
systems from local equilibrium, at least in the FPU model.

\begin{figure}
  \centering
   \includegraphics[width=8.5cm,clip=true]{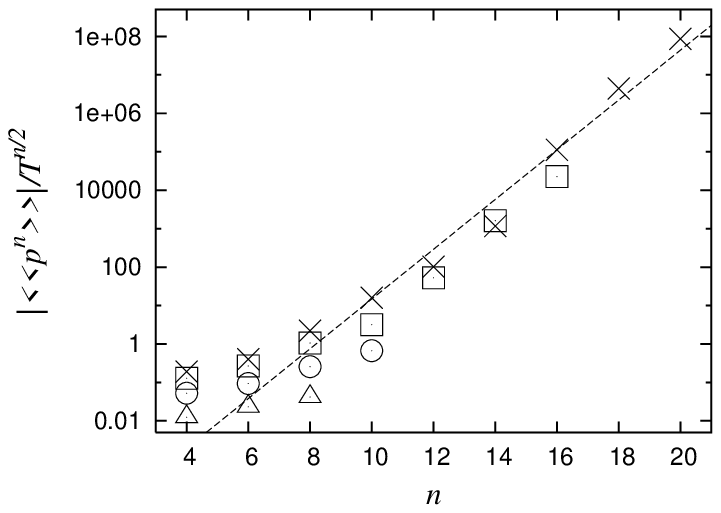}
   \includegraphics[width=8.5cm,clip=true]{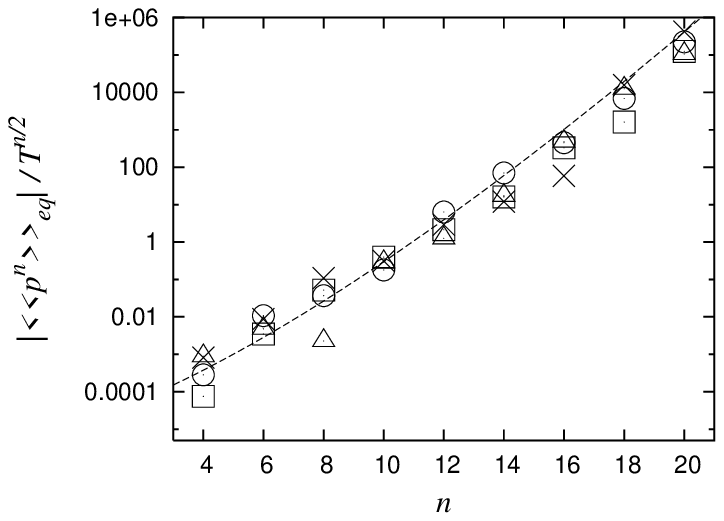}
\caption{(left) Higher order cumulants, $|\cum{p^n}|/T^{n/2} \ (n\leq20)$
  for the four systems in \figno{p2x}, $(T_1^0,T_2^0)=(0.88,16.72), \
  (\times)$,   $(2.4,15.2),\ (\Box)$, $(4.4,13.2),\ (\bigcirc)$ and
  $(6.6,11.0),\ (\triangle)$.  
  Only points with reasonably small error are shown.
  The dashed line is $5.0\times   10^{-6}\exp(1.5n)$ drawn for
  comparison.  (right) The equilibrium cumulants for $L=16$  
  ($\times$), $L=32$ ($\Box$), $L=64$ ($\triangle$)  and $L=128$
  ($\bigcirc$) compared to the 
  rough estimate, Eq.~\eqnn{error2} (dashes). The  cumulants $\cum{p^2}$
  were measured at the middle of the system with number of samples
  $N=10^9$ at $T=8.8$.}
\label{fig:cumHigh}
\end{figure}

It is possible to examine the characteristics of the higher order
cumulants. 
It should be noted that unlike the even moments $\vev{p^{2n}}$, even
cumulants, $\cum{p^{2n}}$, need not be positive and in general will
not be. So to study the general trend of the cumulants for higher
order, we examine the magnitude of the cumulants.  
In \figno{cumHigh}~(left), we show the behavior of the
cumulants up to 20-th order for the same four systems in \figno{p2x},
specifically for the point at which the momentum distributions in
\figno{pDist} were computed.
Only data points with reasonable error  are
shown and an explanation of the relevant errors is given below.
We see an increase in the magnitude with the order is roughly
exponential.  This growth is far milder than the $(2n)!$ seen in
\eqnn{genfn}.

The behavior of the higher order
cumulants is of some import and we briefly explain semi-quantitatively
why they are difficult to obtain. The difficulty lies mainly in the
statistical error in the simulations. This can be  estimated
from the number of  samples for computing the expectation
values as
\begin{equation}
  \label{eq:error1}
  {\Delta\vev{p^{n}}\over\vev{p^{n}}}\sim{n\over\sqrt{N}}
\end{equation}
where $\Delta$ denotes the error and $N$ is the total number of
samples or the number of time steps in the simulation. Note that
$\cum{p^{n}}=\vev{p^n}+\ldots$ so that an error estimate for the
moment should suffice as the error estimate for the cumulant.
An adequate value for the moment can be obtained in equilibrium,
\begin{equation}
  \label{eq:momEq}
  {\vev{p^n}\over T^{n/2}}\sim(n-1)!!
\end{equation}
Combining these relations, we find the statistical error for the
cumulants which increases rapidly for higher order cumulants.
\begin{equation}
  \label{eq:error2}
  \Delta\left(\cum{p^{n}}\over T^{n/2}\right)\sim
  {  (n-1)!! \,n\over\sqrt{N}}
\end{equation}
These estimates for the error also apply to the equilibrium
situation. In contrast to the non-equilibrium cumulants, the
equilibrium cumulants should vanish, with the exception of
$\cum{p^2}$. As the measured values will converge to zero, at any
given time-step in the simulation, their values will be generically
non-zero. 
In \figno{cumHigh}~(right), we compare the {\it equilibrium}
cumulants, in the middle of the system to the above error
estimates. It can be seen that the rough estimate~\eqnn{error2}
seems to be consistent with the results. As one samples more  ($N$
increases), these will tend to zero. However, for a finite sample
size, this is found to explain the order of the uncertainty.

With $N=10^9$ time-steps
---  which we used for the values in \figno{cumHigh} ---  for 8
and 10-th order cumulants, the errors are 0.03 and 0.3. As we can 
see from \figno{cumHigh}~(left), this means that we can obtain  up to
the  8 or 10-th cumulant with reasonable error for the four systems
but the higher order cumulants are expected to be unreliable for
systems closer to equilibrium.
These error estimates are
quite consistent with the estimates we obtain from the statistical
properties of the simulations. These errors can be overcome with
higher statistics which quickly becomes unrealistic for higher
order. We have analyzed systems with various other temperature
boundary conditions and $L$ and have found the increasing behavior of
the cumulants seen in \figno{cumHigh}~(left) to be quite  generic.
\section{Summary and discussions}
\label{sec:summary}
The spatial distribution of cumulants in non-equilibrium steady states
under thermal gradients were predicted from general considerations and
tested in the the FPU model. The understanding of the temperature
profile for a given non-equilibrium steady state, combined with the
deviations of physical observables from their equilibrium values, can
be used to develop a consistent description of the spatial
distribution of observables. In principle, the behavior of
observables probably have higher order corrections in the
non-equilibrium nature of the system, which in this case is $\nabla
T$, but higher order effects could not be separated within the current
numerical simulation results.

We quantitatively analyzed the relation between the momentum cumulants
and the distribution in the non-equilibrium steady state. It was found
that the lower order cumulants characterize the difference of the
non-equilibrium distribution from the one in local equilibrium quite
well. Understanding and characterizing the properties of the
distribution is of manifest importance since the distribution function
for physical variables allows us to compute {\it any} observable
constructed from these variables. To understand the properties of any
local variable in the non-equilibrium state, the physical properties
of the coordinate variables also need to to be clarified.

A comment is perhaps in order: lack of local equilibrium behavior can
in some cases be attributed to the lack of coarse
graining\cite{fluct}. Heuristically speaking,
if one does not have a large
number of degrees of freedom, one cannot see the equilibrium behavior.
This is a {\it different} phenomenon from the case at hand, since the
effective number of degrees of freedom is the number of samples in the
ensemble average which is taken in the time averaging procedure. This
number is huge. In fact, as is well known, in these types of ensembles,
it makes perfect sense to talk even about the statistical
mechanics of one spin degrees of freedom. This is also quite clear
from our results; the deviations from local equilibrium seen in
\figno{p4}, \figno{p6}, Eq.~\eqnn{spatial1} are of definite sign
and no amount of averaging over space will make it zero. So coarse
graining will {\it not} average out the violations of local
equilibrium seen above. Also, the non-local equilibrium properties
found in this paper pertain to systems in the non-equilibrium steady
state and therefore are not transient.

We have also performed similar analyses of spatial distributions on
the $\phi^4$ model. The physical properties of the model are different
from those of FPU model and we intend to report on this in the near
future.

\end{document}